\documentclass{revtex4}
\usepackage{graphicx}
\usepackage{fancyhdr}
\usepackage{amsmath}
\usepackage{color}

\begin{document}

%Title of paper
\title{Vacuum stability of supersymmetric extended Higgs sectors \\ with a discrete symmetry}

\author{Naoki Machida}
\affiliation{Department of Physics, University of Toyama, 3190 Gofuku, Toyama 930-8555, Japan}

\begin{abstract}
We study vacuum stability of supersymmetric extended Higgs sectors with a discrete $Z_2$ symmetry.
These models may be able to explain dark matter, neutrino masses, 
baryon asymmetry of the Universe and hierarchy problem simultaneously at the TeV scale.
This $Z_2$ symmetry could be broken spontaneously.
We examine the stability of the $Z_2$ symmetric vacuum at the tree level 
in several models with such extended Higgs sectors. 
\end{abstract}

\maketitle

\thispagestyle{fancy}

\section{Introduction}
The standard model (SM) is successful in describing the particle physics below the electroweak scale.
This model is consistent with the current data from experiments at the Large Hadron Collider, Tevatron, LEP and so on.
However, there are several well-known problems which cannot be expained in the SM.
%%%%%%%%%%%%%%%%%%%%%%%
There is no candidate for dark matter (DM) \cite{Hinshaw:2012aka}.
The neutrino oscillation data indicate that neutrinos have tiny masses and mix with each other \cite{Beringer:1900zz}.
Furthermore, matter and anti-matter asymmetry in the Universe has been addressed.
%%%%%%%%%%%%%%%%%%%%%%%
In order to solve these problems, we must consider a new physics model beyond the SM.

In this talk, we consider several supersymmetric SMs with extended Higgs sectors,
where a discrete ($Z_2$) symmetry is introduced as a symmetry of the Lagrangian.
These models can be regarded as a supersymmetric extention of radiative seesaw models \cite{Krauss:2002px, Ma:2006km, Aoki:2008av}.
They may be able to solve the problems of DM, neutrino masses, 
baryon asymmetry of the Universe and hierarchy problem at the same time.
The lightest supersymmetric particle (LSP), which is R-parity odd, is then a candidate of the DM.
In addition, the lightest $Z_2$-odd particle is also another candidate of the DM \cite{Aoki:2011he}.
This is a kind of multi-component DM scenarios \cite{Aoki:2012ub}.
Since tree level neutrino Yukawa couplings are forbidden due to the $Z_2$-odd right-handed neutrinos $N_R^i~(i=1$-$3$),
neutrino masses are generated at the loop level.
In these models, baryon asymmetry may be generated at the electroweak phase transition (EWPT) \cite{Aoki:2008av}. 
Needless to say, hierarchy problem can be solved by supersymmetry.

These models predict many $Z_2$-odd scalar particles, 
so that the $Z_2$ symmetry may be broken spontaneously when these scalars get vacuum expectation values VEVs.
We would like to consider the case where the $Z_2$ symmetry is not broken spontaneously.
%%%
There are some previous studies in non-SUSY models \cite{Ginzburg:2010wa, Gonderinger:2009jp}.
%%%
We examine the stability of the $Z_2$ symmetric vacuum in SUSY models, and we find the parameter region which
preserves the $Z_2$ symmetry.

\section{Model}
We focus on supersymmetric extended Higgs sectors with a $Z_2$ symmetry.
We consider the models listed below:
\begin{itemize}
\item 4HDM         \hspace{0.65em}: two $Z_2$-even doublets and two $Z_2$-odd doublets \cite{Aoki:2011yy}.
\item 4HDMS        \hspace{0.15em}: two $Z_2$-even doublets,    two $Z_2$-odd doublets and one $Z_2$-odd neutral singlet \cite{Aoki:2011he}.
\item 4HDM$\Omega$ \hspace{0em}: two $Z_2$-even doublets,    two $Z_2$-odd doublets and two $Z_2$-odd charged singlets 
\cite{Kanemura:2011fy}.
\end{itemize}
In these models, since $Z_2$-even doublets are the identical ones in the minimial supersymmetric SM (MSSM), 
the lightest Higgs boson mass is predicted to be the same as that in the MSSM at the tree level when the $Z_2$ symmetry is unbroken.
Extra $Z_2$-odd Higgs boson loop can enhance the Higgs boson mass at the one loop level.
This one loop contribution can naturally reproduce the mass 126~GeV for the Higgs boson.
In these models, the condition of sufficiently strong 1st order EWPT with $m_h$=126~GeV, which is required for successful
electroweak baryogenesis, leads to relatively strong coupling constants.
In general, such a theory brings the Landau pole at 10-100~{\rm TeV} due to strong coupling constants at the electroweak scale,
so that there should be the cut-off scale below 10-100~TeV, above which a more fundamental theory appears \cite{Harnik:2003rs,Kanemura:2012uy}.
These models may be able to solve problems for DM, neutrino masses, 
baryon asymmetry of the Universe simultaneously at the TeV scale~\cite{Kanemura:2012hr}.
A merit to consider these models is the testability at collider experiments.

First, we consider a simple toy model,
because the vacuum structure in the 4HDM, the 4HDMS and the 4HDM$\Omega$ is quite complicated.
The Higgs sector in the toy model consists of one $Z_2$-even isospin doublet, one $Z_2$-odd isospin doublet 
and one $Z_2$-odd neutral singlet (see TABLE \ref{table:one}).
The superpotential and the soft SUSY breaking terms relevant to the Higgs potential are
\begin{align}
W                 &= \frac{\mu_S}{2} S^2 + \lambda H_1' \cdot H_2 S , \label{eq:Wtoy}\\
- {\cal L}_{soft} &= m_1^2 |H_1'|^2 + m_2^2 |H_2|^2 + m_S^2 |S|^2 + ({\bar{A} H_1' \cdot H_2 S + {\rm h.c.}}). \label{eq:softoy}
\end{align}
The Higgs potential in this toy model is 
\begin{align}
V_{Higgs} &= m_1^2 |H_1'|^2 + m_2^2 |H_2|^2 + M_S^2 |S|^2 + (A H_1' \cdot H_2 S + {\rm h.c.}) \nonumber \\
          &+ \frac{g^2 + g'^2}{8}(|H_2|^2 - |H_1'|^2)^2 + \frac{g^2}{2}|H_1'^\dagger H_2|^2 \nonumber \\
          &+ |\lambda|^2 (|H_1|^2 |S|^2 + |H_2|^2 |S|^2 + |H_1' \cdot H_2|^2), \label{eq:Vtoy}
\end{align}
where $\bar{A} + \lambda \mu_S^* \equiv A$, $M_S^2 \equiv m_S^2 + |\mu_S|^2$ and 
$g~(g'$) is the gauge coupling constant of the $SU(2)_L~(U(1)_Y)$.
Notice that this model is an unrealistic model.
For example, masses of down-type quraks and charged leptons are zero when the $Z_2$ symmetry is unbroken.

Second, we consider the 4HDMS.
This model has four isospin doublets (two of them are $Z_2$-odd) and one $Z_2$-odd neutral singlet (see TABLE \ref{table:tow}).
The superpotential and the soft SUSY breaking terms relevant to the Higgs potential are
\begin{align}
W &= - \mu H_d \cdot H_u + \lambda_d H_d \cdot \Phi_u S + \lambda_u H_u \cdot \Phi_d S,  \label{eq:Ws}\\
- {\cal L}_{soft} &= M_d^2 |H_d|^2 + M_u^2 |H_u|^2 + M_d'^2 |\Phi_d|^2+ M_u'^2 |\Phi_u|^2 + M_S^2 |S|^2 \nonumber \\
                  &+(B \mu H_d \cdot H_u + A_d H_d \cdot \Phi_u S + A_u H_u \cdot \Phi_d S+ {\rm h.c.}). \label{eq:sofs}
\end{align}
In Eqs.~(\ref{eq:Ws}) and (\ref{eq:sofs}), we introduce a global $U(1)$ symmetry for simplicity,
so that the $\mu$-terms and the B-terms in the $Z_2$-odd sector vanish.
The Higgs potential in this 4HDMS is 
\begin{align}
V_{Higgs} &= m_1^2 |H_d^c|^2 + m_2^2 |H_u|^2 
           + M_d'^2 |\Phi_d^c|^2 + M_u'^2 |\Phi_u|^2 + M_S^2 |S|^2 
           + \{m_3^2 H_d^{c \dagger} H_u + {\rm h.c.} \} \nonumber \\
          &- \bigl\{ A_d (H_d^{c \dagger} \Phi_u S) + A_u (H_u^\dagger \Phi_d^c S) 
           + \lambda_u \mu^* (H_d^{\dagger} \Phi_d S)   + \lambda_d \mu^* (H_u^{c \dagger} \Phi_u^c S) + {\rm h.c.} \bigr\} \nonumber \\
          &+ \{ |\lambda_d|^2 |\Phi_u|^2 |S|^2 + |\lambda_u|^2 |\Phi_d^c|^2 |S|^2
           + |\lambda_u|^2 |H_u|^2 |S|^2 + |\lambda_d|^2 |H_d^c|^2 |S|^2 
           + |\lambda_d H_d^{c \dagger} \Phi_u + \lambda_u \Phi_d^{c \dagger} H_u|^2\} \nonumber \\
          &+ \frac{g^2+g'^2}{8} (|H_u|^2+|\Phi_u|^2-|H_d^c|^2-|\Phi_d^c|^2)^2 \nonumber \\
          &+ \frac{g^2}{2} \{   |H_d^c \cdot H_u|^2 
                              + |\Phi_d^c \cdot H_u|^2
                              + |H_d^c \cdot \Phi_u|^2
                              + |\Phi_d^c \cdot \Phi_u|^2
                              - |\Phi_d^c \cdot H_d^c|^2
                              - |\Phi_u \cdot H_u|^2 \}, \label{eq:V4s}
\end{align}
%このヒッグスセクターはSUSY　Ma模型のものと同じ　　　U(1)_gを除いて
where $H_d^c (\Phi_d^c) \equiv (i \sigma_2) H_d^* (\Phi_d^*)$, $m_{1(2)}^2 \equiv M_{d(u)}^2 + |\mu|^2$ and $\ B \mu \equiv m_3^2$.
This Higgs sector corresponds to that of the supersymmetric radiative seesaw model up to $U(1)$ symmetry \cite{Ma:2006uv}.
This model may be able to explain the existence of DM (the lightest $Z_2$-odd and/or R-parity odd particle) 
and neutrino masses at the one loop level.
\\

\begin{center}
\begin{table}[h]
\begin{tabular}{ccc|c}
\hline
       & $SU(2)_L$ & $U(1)_Y$        & $Z_2$ \\ \hline \hline
$H_1'$ & 2         & + $\frac{1}{2}$ &  --   \\
$H_2$  & 2         & --$\frac{1}{2}$ &  +    \\
S      & 1         & 0               &  --   \\
\hline
\end{tabular}
\caption{The Higgs sector in the toy model.}
\label{table:one}
\end{table}

\begin{table}[h]
\begin{tabular}{ccc|cc}
\hline
          & $SU(2)_L$ & $U(1)_Y$        & $U(1)$    & $Z_2$ \\ \hline \hline
$H_d$     & 2         & --$\frac{1}{2}$ & 0        & +     \\
$H_u$     & 2         & + $\frac{1}{2}$ & 0        & +     \\ \hline
$\Phi_d$  & 2         & --$\frac{1}{2}$ & +1       & --    \\
$\Phi_u$  & 2         & + $\frac{1}{2}$ & +1       & --    \\
S         & 1         & 0               & --1      & --    \\
\hline
\end{tabular}
\caption{The Higgs sector in the 4HDMS.}
\label{table:tow}
\end{table}
\end{center}

\section{analysis}
The $Z_2$ symmetry can be broken spontaneously depending on parameters of the Higgs potential.
We would like to find the parameter region where this symmetry is unbroken spontaneously.
We impose conditions that the $Z_2$ symmetric vacuum satisfies 
\begin{align}
\left. \mathrm{det} \Biggl( {\frac{\partial^2 V_{Higgs}}{\partial \varphi_i \partial \varphi_j}} \right|
_{\langle \varphi_{i(j)} \rangle = v_{i(j)}} \Biggr) > 0, \\
V_{{\rm broken} \ Z_2} > V_{{\rm unbroken} \ Z_2}.
\end{align}
The first condition means that the scalar masses are non-tachyonic on the $Z_2$ symmetric vacuum.
The second condition means that the $Z_2$ symmetric vacuum (VEV $v= 246$~GeV) is the global minimum of
the Higgs potential if there are another local minima of the potential where the $Z_2$ symmetry is broken.
We use these two conditions to constrain the parameters of the Lagrangian.
We find that there is the parameter region where the $Z_2$ symmetry is unbroken.

\begin{figure}[h]
\begin{center}
\includegraphics[width=100mm]{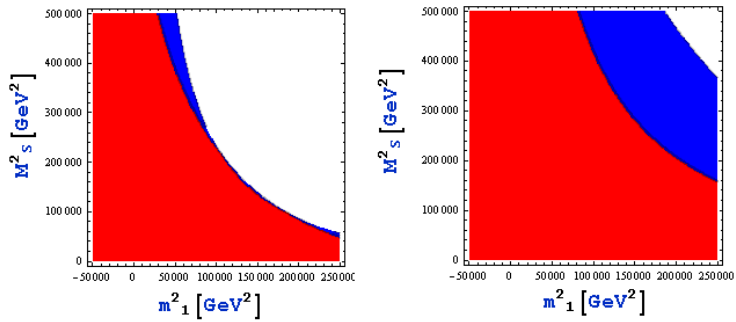}
\end{center}
\caption{The vacuum stability in the toy model. We take $\lambda = 1$. 
$A$ = 700~{\rm GeV} at left figure and $A$ = 1000~{\rm GeV} at right figure.
In the red  region, the $Z_2$ symmetric realistic vacuum with $v=246$~GeV becomes tachyonic.
In the blue region, the realistic vacuum is not a global minimum.}
\label{fig:four}
\end{figure}

In the toy model, there are one $Z_2$-even neutral scalar boson, four $Z_2$-odd neutral scalar bosons (two CP-even and two CP-odd) and 
one $Z_2$-odd charged scalar boson.
Since we focus on spontaneous $Z_2$ symmetry breaking, 
we neglect the direction of charge breaking in the potential analysis.
We consider the case that the $\lambda$ in Eq.~(1) is of order one, 
which is required to realize successful electroweak baryogenesis in the 4HDMS \cite{Kanemura:2012hr}.
We also require this condition in the toy model.
Trilinear scalar terms (i.e. ; $A H_1' \cdot H_2 S$) play important roles in the potential analysis.
Notice that we here assumed that all parameters are real.
We take $A$ = 700~{\rm GeV} and $A$ = 1000~{\rm GeV} with $\lambda$ = 1.
Our numerical results are shown in FIG. \ref{fig:four}.
%%絵の説明
Vertical axis is $M_s^2~[{\rm GeV}^2]$ and horizontal axis is $m_1^2~[{\rm GeV}^2]$.
In the red  region, the $Z_2$ symmetric realistic vacuum with $v=246$~GeV becomes tachyonic.
In the blue region, the realistic vacuum is not a global minimum.
We can see that the realistic vacuum is stable when trilinear scalar term is small and mass parameters are large.
\\

\begin{figure}[h]
\begin{center}
\includegraphics[width=110mm]{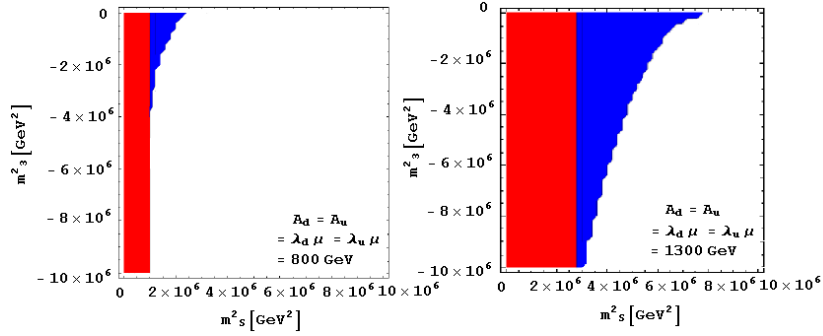}
\end{center}
\caption{The vacuum stability in the 4HDMS. 
We take $\lambda_d=\lambda_u = 1$ and $M_d'=M_u'=300~{\rm GeV}$.
Trilinear scalar terms = 800~GeV at left figure and trilinear scalar terms = 1300~GeV at right figure.
In the red region, the $Z_2$ symmetric realistic vacuum with $v=246$~GeV becomes tachyonic.
In the blue region, the realistic vacuum is not a global minimum.}
\label{fig:one}
\end{figure}

Next we consider the 4HDMS.
There are two $Z_2$-even neutral scalar bosons (they are the MSSM-like scalar bosons at the tree level.)
and three $Z_2$-odd neutral scalar bosons.
The toy model has trilinear scalar terms only from ${\cal L}_{soft}$ in Eq.~(\ref{eq:softoy}).
On the other hand, the 4HDMS has trilinear scalar terms not only from ${\cal L}_{soft}$ but also from the superpotential~(i.e. ; $\lambda_d \mu (H_u^{c \dagger} \Phi_u^c) S$).
In a strongly coupled theory, these terms become very large, so that they aslo play important role.
For simplicity, we take the universal trilinear scalar parameters as $A_d=A_u=\lambda_d \mu=\lambda_u \mu$ in this analysis.
Our numerical results are shown in FIG. \ref{fig:one}.
%%絵の説明
Vertical axis is $m_3^2~[{\rm GeV}^2]$ and horizontal axis is $m_s^2~[{\rm GeV}^2]$.
In the red  region, the $Z_2$ symmetric realistic vacuum with $v=246$~GeV becomes tachyonic.
In the blue region, the realistic vacuum is not a global minimum.
We can see that the realistic vacuum is stable when trilinear scalar terms are small and mass parameters are large.

\section{Conclusion}
We study the vacuum stability in the toy model (one $Z_2$-even doublet, one $Z_2$-odd doublet and one $Z_2$-odd neutral singlet) 
and the 4HDMS (four doublet (two of them are $Z_2$-odd) and one $Z_2$-odd neutral singlet).
The 4HDMS may be able to explain DM, neutrino masses, baryon asymmetry of the Universe and hierarchy problem at the same time.
Especially, we examine the stability of the $Z_2$ symmetric vacuum at the tree level.
We find that the $Z_2$ symmetry is kept unbroken in the parameter region 
where the trilinear scalar terms are small and mass parameters are large.
However, in the strongly coupled theory, trilinear scalar coupling constants $\lambda_i$ in the superpotential are very large,
so that mass parameters are very large to stabilize the the Higgs potential.
In such parameter regions, it would be more difficult to test models
because the deviation in the triple Higgs boson coupling from the SM prediction is more decoupled \cite{Appelquist:1974tg}.
Further studies on details of the vacuum structure in these SUSY extended Higgs sectors are under way.

\begin{acknowledgments}
The author wishes to thank Prof.~S.~Kanemura and Prof.~T.~Shindou for their encouragement and helpful discussions.
\end{acknowledgments}

\end{document}